A white paper for the ESO Expanding Horizons initiative

# Unlocking the physics of dwarf galaxies in the 2040s: The case for a next-generation wide-field spectroscopic facility with fibres and IFUs

Crescenzo Tortora[1], Daniela Carollo[2], Leslie Hunt[3], Francine Marleau[4], Rossella Ragusa[1], Teymoor Saifollahi[5], Fernando Buitrago[6], Michele Cantiello[7], Christopher Conselice[8], Francesco De Paolis[9], Sven De Rijcke[10], Pierre-Alain Duc[5], Anna Gallazzi[3], Pavel E. Mancera Piña[11], Anna Ferre Mateu[12], Garreth Martin[13], Mar Mezcua[14], Nicola R. Napolitano[15], Lucia Pozzetti[16], Justin Read[17], Marina Rejkuba[18], Joanna Sakowska[19], Paolo Salucci[20], Elham Saremi[21], Diana Scognamiglio[22], Francesco Shankar[21], Marilena Spavone[1], and the Euclid Local Universe SWG

*Affiliations:*
[1]INAF – Osservatorio Astronomico di Capodimonte, Salita Moiariello 16, I-80131 Napoli, Italy
[2]INAF – Osservatorio Astronomico di Trieste, Via G. B. Tiepolo 11, 34143 Trieste, Italy
[3]INAF – Osservatorio Astrofisico di Arcetri, Largo Enrico Fermi 5, 50125 Firenze, Italy
[4]Universität Innsbruck, Institut für Astro- und Teilchenphysik, Technikerstr. 25/8, 6020 Innsbruck, Austria
[5]Université de Strasbourg, CNRS, Observatoire astronomique de Strasbourg, UMR 7550, 67000 Strasbourg, France
[6]Departamento de Física Teórica, Atómica y Óptica, Universidad de Valladolid, 47011 Valladolid, Spain; Instituto de Astrofísica e Ciências do Espaço, Faculdade de Ciências, Universidade de Lisboa, Tapada da Ajuda, 1349-018 Lisboa, Portugal
[7]INAF – Osservatorio Astronomico d'Abruzzo, Via Maggini, 64100 Teramo, Italy
[8]Jodrell Bank Centre for Astrophysics, Department of Physics and Astronomy, University of Manchester, Oxford Road, Manchester M13 9PL, UK
[9]Department of Mathematics and Physics "Ennio De Giorgi", University of Salento, Lecce, Italy
[10]Ghent University, Department of Physics & Astronomy, Krijgslaan 281, S9, B-9000 Ghent, Belgium
[11]Leiden Observatory, Leiden University, Einsteinweg 55, 2333 CC Leiden, The Netherlands
[12]Instituto de Astrofísica de Canarias, Vía Láctea, 38205 La Laguna, Tenerife, Spain
[13]School of Physics and Astronomy, University of Nottingham, University Park, Nottingham NG7 2RD, UK
[14]Institute of Space Sciences (ICE, CSIC), Campus UAB, Carrer de Can Magrans s/n, 08193 Barcelona, Spain
[15]Department of Physics "E. Pancini", University of Naples Federico II, Via Cintia 21, 80126 Naples, Italy
[16]INAF – Osservatorio di Astrofisica e Scienza dello Spazio di Bologna, Via Gobetti 93/3, I-40129 Bologna, Italy
[17]Department of Physics, University of Surrey, Stag Hill Campus, Guildford GU2 7XH, UK
[18]European Southern Observatory, 85748 Garching bei München, Germany
[19]Instituto de Astrofísica de Andalucía (CSIC), Glorieta de la Astronomía, E-18080 Granada, Spain
[20]SISSA, Via Bonomea 265, 34100 Trieste, Italy
[21]School of Physics & Astronomy, University of Southampton, Highfield Campus, Southampton SO17 1BJ, UK
[22]Jet Propulsion Laboratory, California Institute of Technology, 4800 Oak Grove Drive, Pasadena, CA 91109, USA

# Context and Open Questions for the 2040s

Dwarf galaxies ($M_* \lesssim 10^9$ $M_\odot$) are the most numerous galaxies in the Universe (e.g., McConnachie 2012, AJ, 144, 4), yet they remain among the least understood. Their shallow potentials make them uniquely sensitive to baryonic feedback (Dekel & Silk 1986, ApJ, 303, 39), tidal interactions (Mayer et al. 2001, ApJ, 559, 754), and the microphysics of dark matter (DM, Bullock & Boylan-Kolchin 2017, ARA&A, 55, 343). Their low luminosities ($M_r > -18$) and diffuse morphologies make them excellent tracers of the interplay between baryonic and DM processes (Pontzen & Governato 2012, MNRAS, 421, 3464). Moreover, dwarf galaxies play a central role in testing cosmological models. The so-called "missing satellites problem", namely the discrepancy between the large number of predicted DM subhalos in ΛCDM simulations and the comparatively sparse observed dwarf population around the Milky Way (MW), remains an open issue despite the discovery of numerous ultra-faint dwarfs (Bullock & Boylan-Kolchin 2017). The "too-big-to-fail problem" refers to the difficulty of simultaneously explaining the observed kinematics of bright dwarf galaxies and their number density within the ΛCDM framework (Papastergis & Shankar 2016, A&A 591, A58). N-body simulations predict centrally concentrated, "cuspy" DM profiles in dwarf halos (Moore et al. 1999, MNRAS, 310, 1147), while observations of Local Group (LG) dwarfs, and of systems beyond the LG, often favor shallower, "cored" profiles. Including baryons and feedback can reconcile this issue (Read et al. 2016, MNRAS, 459, 2573), although accurately modeling these processes remains challenging, and opens the possibility for alternative DM flavours (Bullock & Boylan-Kolchin 2017). The existence of planar and corotating structures, such as the Vast Polar Structure around the MW (Pawlowski et al. 2012, MNRAS, 423, 1109), the Great Plane of Andromeda (Ibata et al. 2013, Nature, 493, 62), and Cen A (Müller et al. 2018, Sci, 359, 534) further challenges cosmological expectations, as such satellite distributions are rare in ΛCDM simulations (Pawlowski 2014, MNRAS, 442, 2362; Cautun et al. 2015, MNRAS, 452, 3838). Together, these small-scale tensions underscore the need for comprehensive observational constraints on dwarf galaxy kinematics, stellar populations, and DM content across a wide range of environments.

While the LG has been the primary laboratory for dwarf galaxy studies, it remains unclear whether these small-scale tensions are universal. Extending dwarf galaxy surveys to nearby galaxy groups and isolated hosts in the Local Volume ($z < 0.2$) is therefore critical (Müller et al. 2015, A&A, 583, 79; 2018, A&A, 615, 105). Modern wide-field, high spatial resolution, deep and multi band imaging has enabled the detection of faint, unresolved systems down to surface brightness (SB) levels of $\mu_r \sim$ 29–30 mag arcsec$^{-2}$, allowing the discovery of dwarfs with luminosities comparable to classical LG satellites ($M_r \sim -6$) at distances of 4–5 Mpc and expanding the census of dwarfs in nearby clusters and groups. The Fornax Deep Survey (FDS; Venhola et al. 2018, A&A, 620, 165) and the Next Generation Fornax Survey (NGFS; Muñoz et al. 2015, ApJ, 813L, 15) have catalogued hundreds of dwarfs down to low SB levels, characterizing their properties, nucleation fraction, and spatial distribution. Similar efforts in the Virgo cluster (VEGAS: Capaccioli et al. 2015, A&A, 581, A10) have revealed numerous dwarfs and ultra-diffuse galaxies, highlighting the role of the environment in shaping their properties. Surveys such as MATLAS (Habas et al. 2020, MNRAS, 491, 1901; Marleau et al. 2021, A&A, 654, 105) and UNIONS (Heesters et al. 2025, A&A, 699, 232) have mapped lower-density fields and galaxy groups, uncovering a rich dwarf population beyond clusters. In the LG and nearby groups, systematic searches continue to uncover new dwarfs in the Centaurus group (Müller et al. 2015), around Cen A and NGC 253 (Crnojević et al. 2014, ApJL, 795, L35; 2016, ApJ, 823, 19), and in the M81 and M101 groups (Chiboucas et al. 2009, AJ, 137, 3009; Javanmardi et al. 2016, A&A, 588, A89).

The Euclid mission (Mellier et al. 2025, A&A, 697, 1), observing 14k sq. deg. will open a new window for dwarf studies at larger distances. Early results already demonstrate this capability with over a thousand dwarf candidates in the Perseus cluster (Marleau et al. 2025a, A&A, 697, 12) and ~2700 dwarfs across different environments, within only ~14.25 sq. deg (Marleau et al. 2025b,

arXiv:2503.15335). Complementary imaging from LSST@Rubin (Ivezić et al. 2019, ApJ, 873, 111) will further expand the discovery space for dwarfs. Building on Marleau et al. (2025b), **we anticipate that by the end of the Euclid Wide Survey the number of candidate dwarf galaxies will exceed 2 million** (~190 per sq. deg.), with photometry and structural parameters measured across multiple bands. These surveys will also characterize the properties of several million low-mass (M ~ $10^9$ M$_\odot$) galaxies across cosmic time (z < 1.5; Enia et al. 2025, arXiv:2503.15314), currently even less explored.

## Why a new facility is needed

Despite this progress, we may be unable to follow up on this vast amount of data, as a key limitation remains: spectroscopy. Redshift and distance measurements, internal kinematics (velocity dispersions σ, rotation), chemical abundances, and star formation (SF) histories will be missing for the majority of dwarfs beyond the LG. These data are critical to address fundamental scientific objectives:

- **ΛCDM framework** and the **nature of DM**, including potential signatures of warm, self-interacting, or ultra-light axion DM (Lovell et al. 2014, MNRAS, 439, 300; Bullock & Boylan-Kolchin 2017), can be probed through dwarf galaxies because both their **abundance** and **inner DM density profiles** (e.g., cusps vs. cores) depend sensitively on the underlying DM flavour.
- The **inner DM profiles** of dwarfs and the impact of initial collapse, Supernovae and AGN feedback, reionization, environment and the very first stars (Pontzen & Governato 2012; Read et al. 2016, Verbeke et al. 2015, ApJ, 815, 85).
- The **connection between satellites and their host galaxies**, including abundance (probing the faint end of the luminosity function), spatial distribution, and planar structures (Koposov et al. 2008, ApJ, 686, 279; Pawlowski et al. 2012).
- The integrated and spatially-resolved **stellar populations**, including age-metallicity distributions, SF bursts, and chemical enrichment (Romero-Gómez et al. 2024, MNRAS, 527, 9715).
- **Environmental effects** on SF, quenching, and tidal transformation.
- While imaging data can identify **globular cluster (GC)** candidates (≲20 Mpc) and map their projected spatial distributions, spectroscopic follow-up is essential to confirm GC membership and constrain the relation between DM halo mass and the number of GCs, as well as measure their velocity dispersion, which is used as an independent dynamical tracer of the host galaxy's mass (e.g., Pota et al. 2013, MNRAS, 428, 389). It further enables the determination of ages, metallicities, and chemical abundances of GCs, quantities that remain poorly explored outside the LG. Such measurements would allow us to directly link the SF histories of GCs and their host dwarfs, providing a powerful probe of dwarf galaxy evolution and their DM halos.

Addressing these open questions will require wide-field, high-multiplex spectroscopy with sufficient spectral resolution capable of reaching low-SB dwarfs over large areas, as well as deployable IFU observations for spatially resolved kinematics on specific objects. Without such facilities, the unprecedented datasets from Euclid and LSST risk leaving the physics of dwarf galaxies largely unconstrained and predictions of ΛCDM and alternative DM models untestable beyond the LG. While imaging will provide exquisite morphologies and galaxy counts, it will deliver only weak constraints on SF histories, and essentially no constraints on distances, internal dynamics, dynamical masses, or the microphysics of DM. Dwarfs, where galaxy evolution processes are most visible, are even key to understanding galaxy formation and evolution across all mass scales.

## Why existing 2030s facilities cannot meet the need

Current facilities/surveys such as 4MOST, DESI, WEAVE, PFS, AAOmega cannot reach the necessary depth or multiplexing required for dwarfs with $I_E$≳22–23 mag. For 4-m class telescopes, exposure times of 100–150 hours would be required to measure redshifts and velocity dispersions, whereas 8-m telescopes would reduce this by roughly a factor of four. Euclid spectroscopy is not optimized for low-SB extended sources. The ELT, with its 39-m aperture, could achieve these measurements in only 1–1.5 hours; however, it lacks the multiplexing capability required to build statistically significant samples over large areas within reasonable observing times. By 2035-2040, >1 million dwarf candidates in the southern hemisphere will lack spectroscopy. Closing this gap requires survey-speed scaling that only a large-aperture (≥20 m), wide-field, highly multiplexed facility—currently not existing nor planned—can provide, delivering S/N > 10–15 at $I_E \gtrsim$ 22–23 mag for statistically meaningful samples.

### Technological Requirements

Measuring the internal kinematics, stellar populations, and gas properties of large dwarf samples requires **high S/N spectroscopy** down to $I_E \gtrsim$ 22–23 mag, necessitating a dedicated **large-aperture telescope (>20m)**. The ~1 million dwarfs, together with the much larger population of faint galaxies across redshifts, over 7000 deg² require a **wide-field (~1-3 sq. deg.), high-multiplex, multi-object spectrograph (MOS) capable of thousands to tens of thousands of fibers.**

**Spectral coverage from ~3700 to 10000** Å is required to enable:
(a) stellar population diagnostics (ages, metallicities, abundance ratios);
(b) gas-phase properties (metallicity, SF rate, ionization state, outflows) and AGN diagnostics;
(c) integrated stellar and gas kinematics (velocity dispersions $\sigma \lesssim 20$ km s$^{-1}$).

**Low-resolution spectroscopy (R~3000–4000)** is sufficient for the brightest dwarfs, whereas fainter systems require **medium resolution (R~5000–8000)** to resolve low σ, separate stellar continuum from emission lines, and detect subtle kinematic features such as tidal distortions or central cores.

**Deployable or monolithic IFUs** covering fields of tens of arcseconds to a few arcminutes with sub-arcsecond spatial sampling are essential to map:
a) Spatially-resolved stellar and gas kinematics (velocity dispersion profiles, rotation curves),
b) Gradients in stellar populations (e.g. metallicity) and ionization state, which spectroscopy can uniquely measure to trace chemical enrichment, SF patterns, and feedback across dwarf galaxies.
c) Tidal features and kinematically decoupled components, for selected subsamples of dwarfs.

Therefore, only **a dedicated large-aperture telescope (≥ 20 m) with a wide-field, high-multiplex MOS and deployable or monolithic IFUs** would uniquely enable wide-field spectroscopic surveys for the follow-up of faint sources, providing a **comprehensive physical characterization of dwarf galaxies**. Such capabilities would allow transformative studies of DM cores, baryonic feedback, tidal effects, and environmental dependencies in the nearby Universe, while extending the spectroscopic exploration of the least massive galaxies up to $z \simeq 1.5$ and stringently testing cosmological models. *In the absence of such a facility, upcoming wide-field imaging surveys will vastly outpace spectroscopic capabilities, leaving the physical interpretation of the dwarf galaxy population—and its implications for galaxy formation, DM physics, and cosmology—fundamentally incomplete well into the 2040s.*

## Broader impact

Such a facility would benefit several other science cases: a) scaling relations involving stellar populations and DM content for galaxy populations from low- to high-masses, across redshift and environments; b) validation and characterization of galaxy-scale and cluster-scale gravitational lenses; c) the impact of the cosmic web; d) LSST/Euclid photo-z calibration at the faint end; e) improved cosmological parameter inference from large-scale structure, weak lensing, and galaxy–halo connection models.